# Filament Power Control for Gyrotrons in EAST ECRH System

Weiye Xu*, Handong Xu*, Fukun Liu, Xinsheng Yan, Dajun Wu

*Corresponding authors:* Weiye Xu (e-mail: xuweiye@ipp.cas.cn), Handong Xu (e-mail:xhd@ipp.cas.cn)

The authors are with the Institute of Plasma Physics, Chinese Academy of Sciences, 230031 Hefei, Anhui, China.

Xinsheng Yan is also with the University of Science and Technology of China, 230026 Hefei, Anhui, China.

***Abstract -*** The electron cyclotron heating system including four gyrotrons is under developed in the Institute of Plasma Physics, Chinese Academy of Sciences. The filament is an important part of the gyrotron, which is used to heat the cathode. Each gyrotron has a filament power supply to adjust the filament power. The filament power control and measurement system was developed to control the beam current and the output power of the gyrotrons. The control program was developed in Labview. The filament characteristics were tested using the power control system. The test results show that the filament current can be fitted with the e-exponential function of the filament voltage. It can be seen that as the voltage increases, the filament resistance gradually increases. The 'burst' function was developed to prevent the beam current from dropping too fast. The filament control system can work stably for the gyrotrons.

1. Introduction

The magnetic confinement nuclear fusion is a very promising new energy source. The tokamak [1] is one of the several types of magnetic confinement devices being developed to produce controlled fusion power and it is the leading candidate for a practical fusion reactor. The EAST (Experimental Advanced Superconducting Tokamak) in ASIPP (Institute of Plasma Physics, Chinese Academy of Sciences) is an important tokamak device in China. EAST is the first tokamak to employ superconducting toroidal and poloidal magnets in the world.

Electron cyclotron resonance heating (ECRH) is an important plasma heating method. A 140GHz electron cyclotron resonance heating system for EAST is under developed [2]. It is designed to inject 140GHz/4MW/100s (up to 4MW/1000s in the best of circumstances) wave power to EAST. The ECRH system includes four gyrotrons [3] which are capable of producing 900~1000 kW of RF output power for pulse lengths up to 1000 seconds. So far, three sets of gyrotron systems have been developed in ASIPP. The fourth one is now being manufactured in CPI. We named the gyrotron as #1, #2, #3, and #4 respectively according to the time that the gyrotron was transported to ASIPP. Among them, #1 and #3 gyrotrons are from GYCOM, #2 and #4 gyrotrons are from CPI.

Up to now, the first three gyrotrons have been tested in ASIPP. In recent experimental campaign,

the #1 gyrotron oscillation of 980kW/1s, 903kW/10s, 834kW/95s and 650kW/754s were demonstrated, the #2 gyrotron oscillation of 721kW/0.5s, 647kW/2s, 499kW/80s and 406kW/98s were demonstrated, and the #3 gyrotron oscillation of 787kW/20s, 637kW/100s and 559kW/1000s were demonstrated. The output power is measured using calorimetric method [4].

The electron gun is one of the most important parts of the gyrotron. The filament is an important part of the electron gun and is used to heat the cathode. By controlling the filament power, we can change the cathode temperature to control the beam current, and thus control the gyrotron output power [5, 6].

In this paper, the filament power control system for the four gyrotrons in the EAST ECRH system was developed and the test results of the filament were described in detail. That is very important in the gyrotron test and in the tokamak experiment.

## 2. Filament power supplies for gyrotrons

The filament is used to heat the gyrotron cathode to the required temperature. The DC or AC power supplies can both be used as the filament power supplies for all four gyrotrons. The maximum power of the four gyrotron filaments is different, which is shown in Table 1. The typical power of the GYCOM gyrotron filaments (#1 and #3) is between 1000 W and 1300 W. The CPI gyrotron filaments (#2 and #4) require less power than the GYCOM gyrotron filaments, typically between 200 W and 250 W. The resistance of the four gyrotron filaments is also different, which is shown in Table 1. The resistance of #2 and #4 gyrotron filaments is much larger than that of the #1 and #3 gyrotron filaments.

Table 1 The characteristics of the gyrotron filaments.

| Gyrotron serial number | #1 | #2 | #3 | #4 |
|---|---|---|---|---|
| Maximum filament power, W | 1500 | 250 | 1500 | 250 |
| Cold resistance, $\Omega$ | 0.12 | 1.00 | 0.12 | 1.00 |
| Operating hot resistance, $\Omega$ | 0.97 | 4.29 | 0.97 | 4.29 |
| Filament power type | DC | 50Hz AC | 50Hz AC | 50Hz AC |

### 2.1. Filament power supply for #1 gyrotron

A DC power supply (model No. CFPS 30/50) was used as the filament power for #1 gyrotron. The structure of the filament power supply and its connection to the filament is shown in Fig. 1.

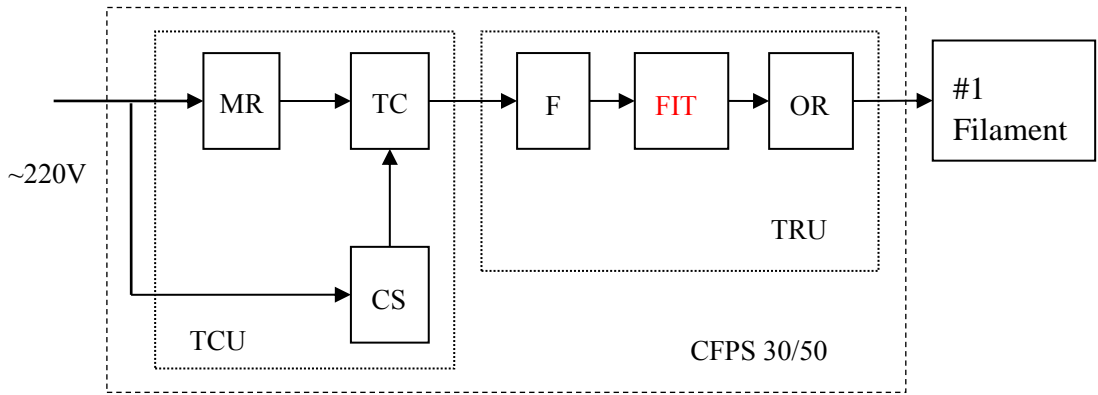

Fig. 1. #1 filament power supply structure and its connection to the filament.

The CFPS 30/50 power supply consists of two units: a Transistor Converter Unit (TCU) and a Transformer Rectifier Unit (TRU).

The TCU module consists of an uncontrolled mains rectifier (MR), a half-bridge transistor converter (TC) with metering capacitors, and a control system (CS). The TC module provides the conversion of the DC voltage from the MR output to the high-frequency alternating voltage. Changing the frequency of the TC module leads to a change in output power of the CFPS 30/50.

The TRU module consists of a passive filter (F), a filament isolation transformer (FIT), and the output rectifier (OR) with a filter. The passive filter is placed between TC and FIT, which protects the TC elements from voltage surges in the heating circuit.

Because the filament itself is connected with the cathode (the cathode voltage is up to -46 kV for #1 gyrotron) directly, the filament isolation transformer must be used to isolate the filament power supply from the filament. The FIT converts the AC voltage on the low voltage side into the AC voltage on the high voltage side. The maximum isolation voltage of FIT is 55 kV. The secondary winding of FIT is loaded on an uncontrolled rectifier with a capacitive filter, which converts the low voltage alternating voltage to a DC heating voltage. The FIT module and the OR module are placed in a metal tank filled with transformer oil, which is the insulating medium.

*2.2. Filament power supply for #2-#4 gyrotrons*

The AC power supplies were chosen to be used as the filament power for #2, #3, and #4 gyrotrons. As shown in Table 1, the power required for all four filaments will not exceed 1500 W. We chose the AC6803A power supplies from Keysight as the filament power supplies for #2 - #4 gyrotrons. Since the filament floats on the cathode high voltage (up to -60 kV for #2 - #4 gyrotrons), a high voltage isolation transformer is required between the filament power supply and the filament [7]. The isolation voltage of the transformer is 100 kV. In order to moderately reduce the current on the low-voltage side, the isolator transformer design ratio is 220:64. The maximum output current of the transformer is 60A. The shield was used between the primary and secondary of the transformer. The insulating material of the transformer is epoxy resin. The connection between the AC 6803A filament power supplies and the gyrotron filaments is shown in Fig. 2.

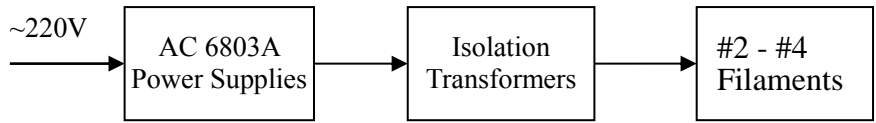

Fig. 2. The connection between the AC 6803A filament power supplies and the gyrotron filaments.

The maximum output power of AC 6803A is 2000 W. It can set as the DC output mode or the AC output mode. The AC output mode with a 50 Hz frequency was used as the filament power supply. In the AC output mode, the voltage ranges can be set as 135 Vrms/270 Vrms (low/high range), the corresponding maximum output RMS current is 20 A/10 A (low/high range).

## 3. Filament power supplies' control system

The filament power affects the gyrotron operation and output RF (Radio Frequency) power. In order to easily set the parameters of the filament power supplies, we developed a filament power supplies' control system. The block diagram of the filament power control system is shown in Fig. 3. The host computer is placed in the control room. The filament power supplies are placed in the heating devices room. The filament power supplies are interlocked by the PLC (Programmable Logic Controller) so that it turns off and remains off (by cutting off the 220 V power) unless the ion pump power supplies are operating, the oil-pump system supplying cooling oil for the gun is working, and the water cooling system is running normally. The 6803A power supplies can be controlled via TCP/IP protocol. The CFPS 30/50 power supply adopts Advantech's ADAM 4022T controller as the control module. The host computer can directly communicate with ADAM 4022T through RS485. Considering the difficulty of wiring between the control room and the heating devices room, and the need for many signals to be transmitted between the control room and the heating devices room, we have placed a single optical fiber as a control signal path between the control room and the heating devices room [8]. We use a RS485-TCP converter to convert RS485 signals to TCP signals for CFPS 30/50 filament power control.

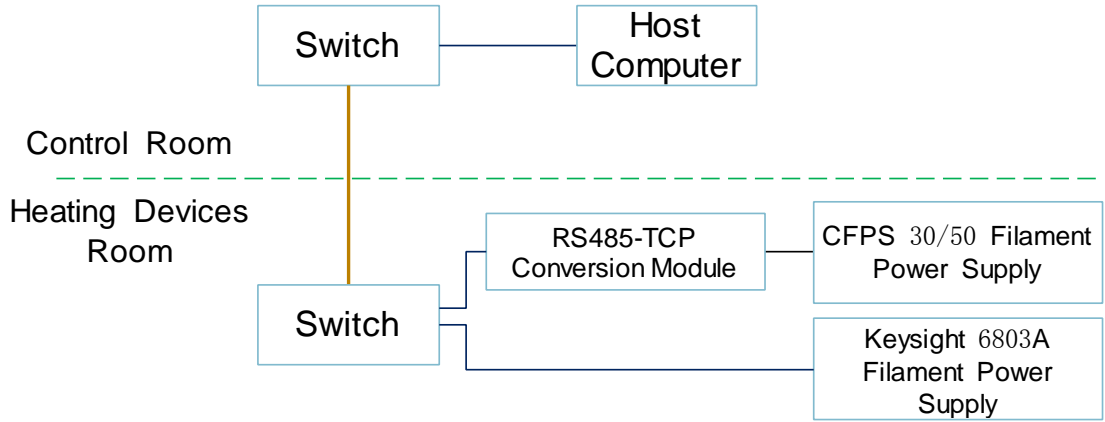

Fig. 3. The block diagram of the filament power control system. Two switches connected by the optical fiber.

### 3.1. CFPS 30/50 filament power supply control

The Advantech's ADAM 4022T controller was used as the control module for the CFPS 30/50 power supply. A RS485-TCP converter is used to control the CFPS 30/50 filament power supply. After the system is connected, the host computer can communicate with the ADAM 4022T module inside the CFPS 30/50 power supply through the virtual serial port. The communication mode settings are shown in Fig. 4. The Advantech ASCII protocol is chosen. For instance, we can set the CFPS power supply 'ON' by writing '#011001' to DO0 port.

The analog inputs and analog outputs are set to a voltage of 0~10V. Since we only use ADAM 4022T as a communication interface and do not need feedback control, the PID control modes of PID 0 and PID 1 need to be set to free mode, which is shown in Fig. 5.

Fig. 4. Communication mode settings for ADAM 4022T

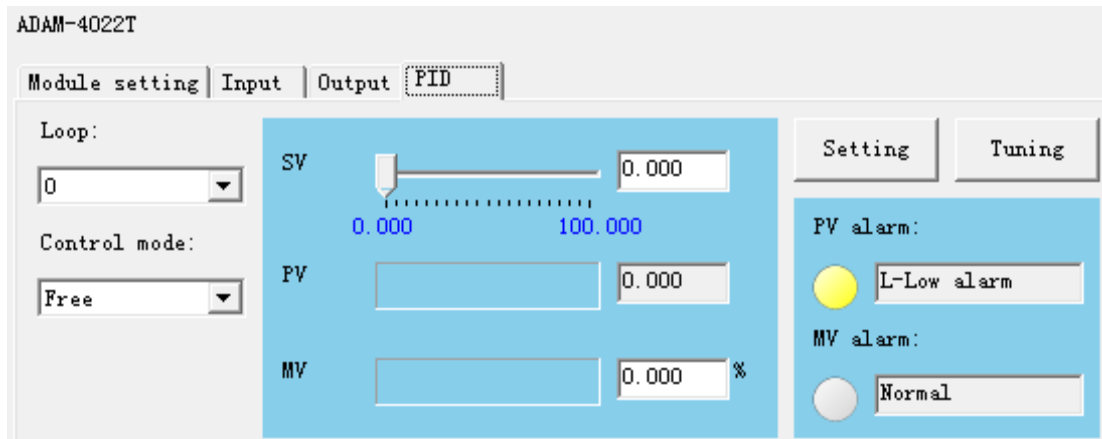

Fig. 5. PID settings for ADAM 4022T

The control program is written in Labview and the program block diagram is shown in Fig. 6. Because the length of the return string of different commands is different, in the initialization program, the terminator '0x0D' is enabled so that the reading operation is ended after the terminator is detected.

The main program shown in Fig. 6 is an 'Event' structure under the While Loop. When the button on the front panel is pressed, the corresponding event will be executed. The program responds with nine events: time-out, 'ON' value changed, 'OFF' value changed, 'Set PF' value changed, 'Up button' value changed, 'Down button' value changed, 'Lock button' value changed, 'Unlock button' value changed, and 'Exit program' value changed. The time-out terminal of the event structure is set to 500 ms, that is, no key on the front panel is pressed for more than 500 ms, the time-out event will be executed. The time-out event program is designed to read the 'ON/OFF' status, read the primary current, and read the primary power value, etc. Since the serial port return value is a string format, we use the 'fraction/index string to numeric conversion' function to convert the string to a double-precision floating-point number and multiply by the corresponding coefficient (power factor 160; current factor 30). In this way, when the current panel button is not pressed, the current power 'ON/OFF' status, current value, power value, etc. are automatically read once every 500 ms. If we use the parallel While Loop structure instead of the time-out event structure to read the power status, the program can not work normally. In the parallel While Loop structure, while the current value is reading, both two commands will be sent simultaneously to the filament power supply if a key on the front panel is pressed. The command appears confusing and the program cannot operate normally. The time-out event structure does not cause the problem of sending multiple commands at the same time. The program runs stably and reliably.

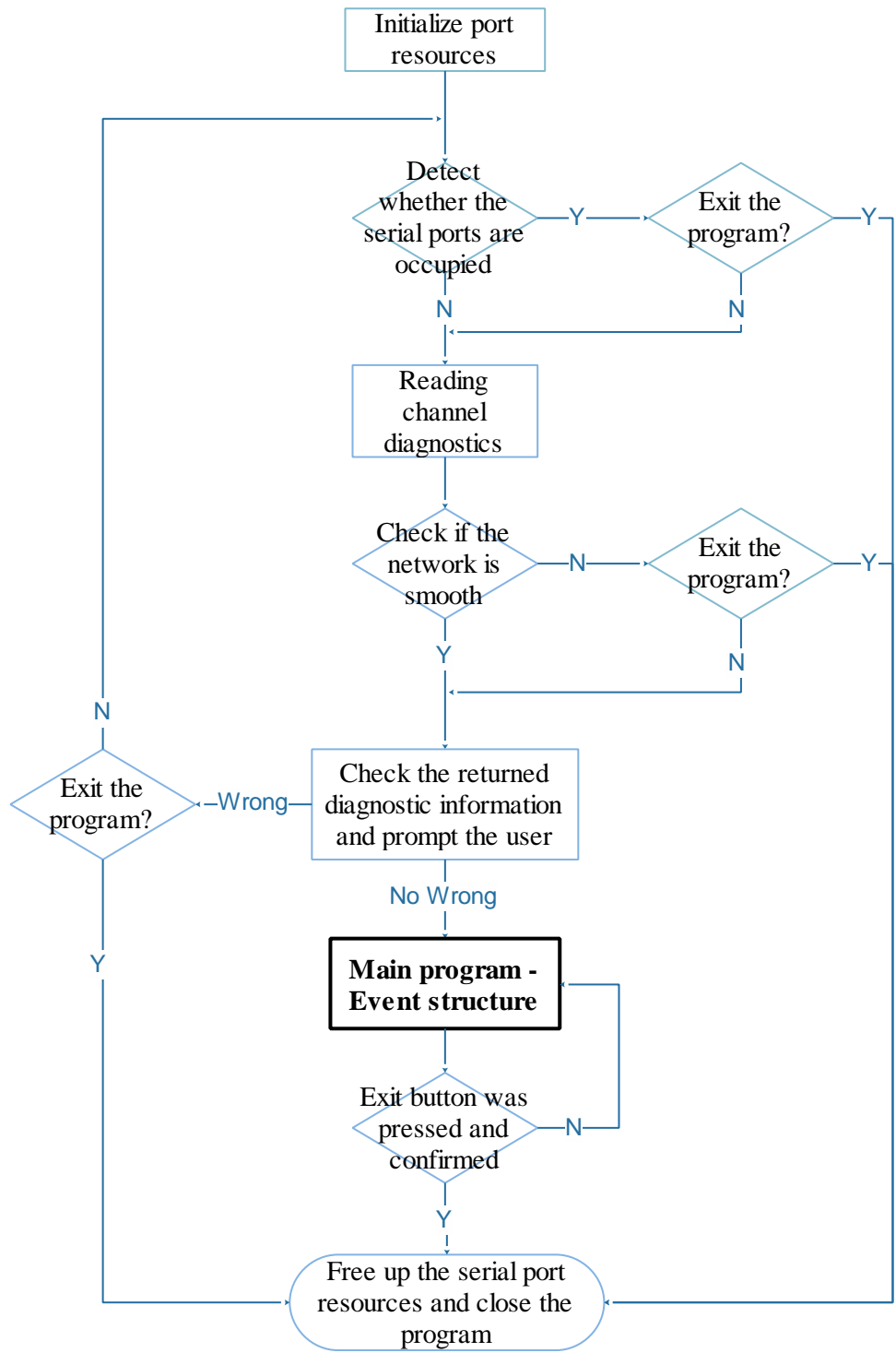

Fig. 6. The program block diagram for controlling the CPFS 30/50 power supply.

The 'Up button' value changed event is used to execute the 'up button pressed event' program. When the up button is pressed, the value of the 'Set PF' control is automatically increased by 0.005 V and the new value is written to the filament power supply. The value is automatically increased by 0.005 V through the 'value' attribute node and the shift register. For security purposes, the PF control value range is set to 0~9.8 V. When this range is exceeded, the control value will not continue to change. The program interface is shown in Fig. 7.

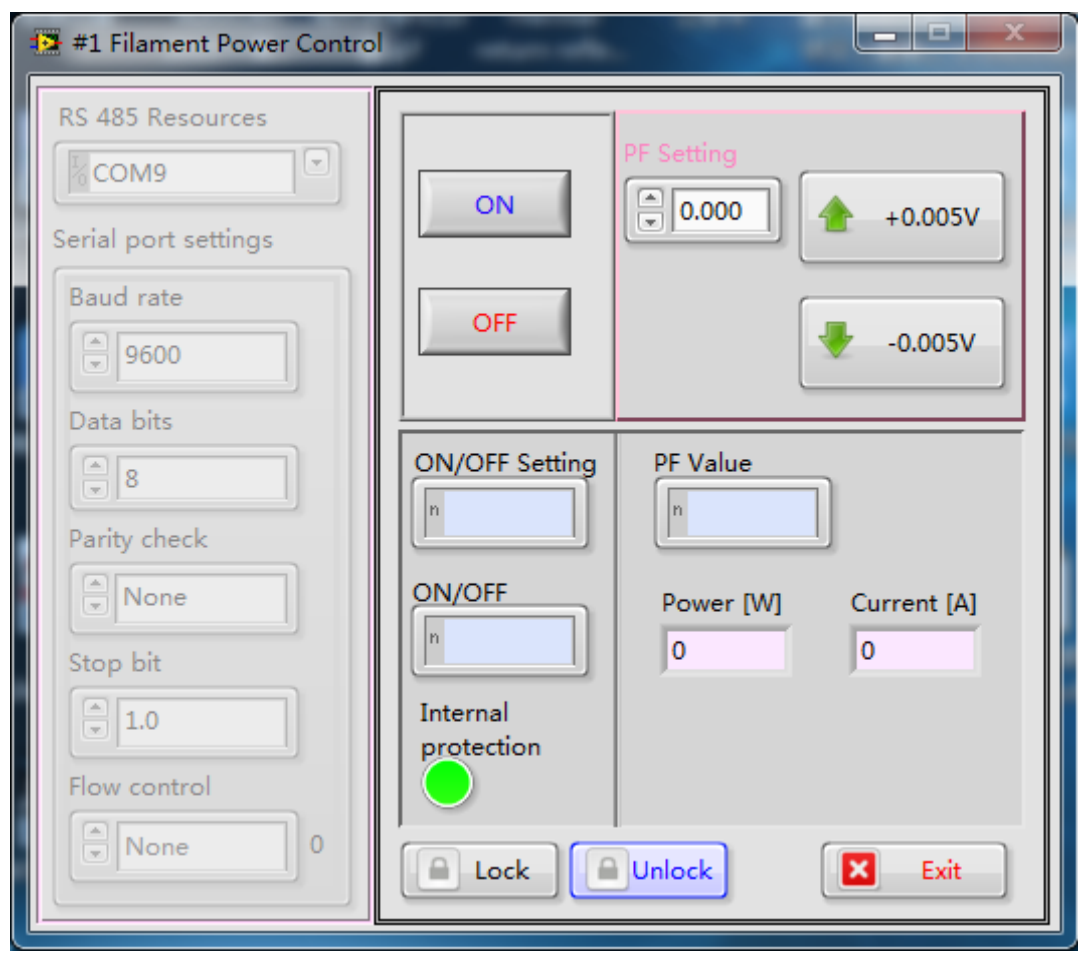

Fig. 7. The program interface for controlling the CPFS 30/50 power supply.

### 3.2. Keysight 6803A filament power supply control

The host computer communicates with the Keysight 6803A power supplies via the TCP/IP protocol. The control language used by the Keysight 6803A power supplies is SCPI (Standard Commands for Programmable Instruments), which is an ASCII-based instrument test measurement control language. The Labview driver for the Keysight 6800 series power supplies based on the SCPI command can be downloaded from the National Instruments official website. We modified the driver program to write the control program for the #2-#4 filament power supplies according to the format and function of the specific command.

The architecture of the Keysight 6803A filament power control program is similar to that of the CFPS 30/50 filament power control program. It first initializes the TCP/IP resources and then enters the event structure under the While Loop. In order to maintain the security of the filament control and make the remote control function not affect the local operation, in the initialization process, unless the forced reset is selected, the program does not clear the power supply state and does not clear the power supply register value after the power is turned on.

The main body of the control program is a 'producer/customer architecture', which is shown in Fig. 8. There are two parallel while loops, one of which is the 'producer' and the other is the 'customer'. The 'producer' loop embedded an event structure is used to monitor the button press events and write the corresponding data (enumeration type data) into the queue. The 'customer' loop embedded an case structure is used to do the corresponding processing according to the received enumeration data.

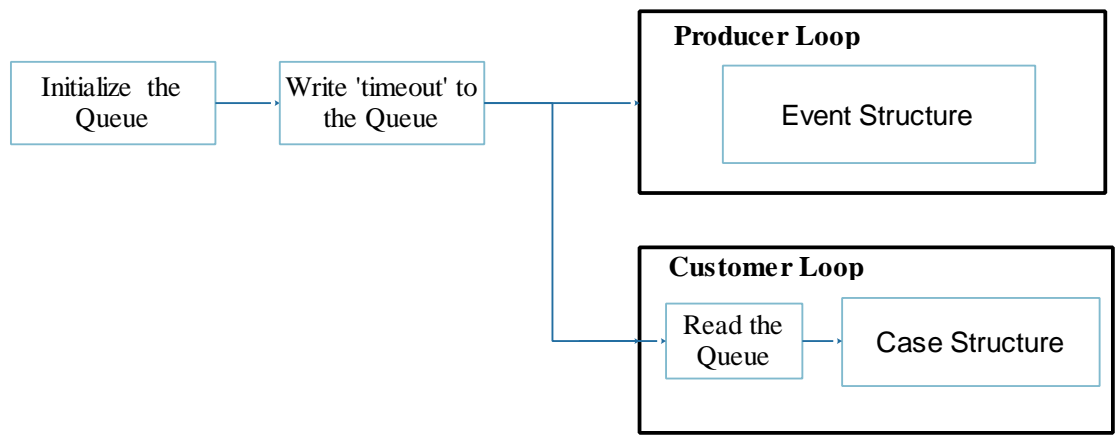

Fig. 8. The program block diagram for controlling the Keysight 6803A power supply.

The program includes seven event responses: time-out, 'Exit program' value changed, 'Set and read filament power parameters' value changed, 'Voltage level' value changed, 'Up button' value changed, 'Down button' value changed, 'Burst' value changed. When the value of the corresponding control on the current panel changes, the corresponding event is triggered. In order to ensure that the right instrument is connected, all VISA (Virtual Instrument Software Architecture) resources used in the event structure are from the initialization program, but not from the value attribute nodes of the VISA control or local variables of the VISA control. Similar to the CFPS 30/50 control program, the 'time-out event' is used to periodically (500ms) update the power supply status. For the safe operation of the gyrotron filament, all voltage source setting parameters are set only if the 'Set parameters' button is pressed. The ON/OFF state will be automatically detected before writing the parameter. If the power is off when writing, normal writing will be performed. If the power is on when writing the settings, the settings will no longer be written directly and the dialog box will pop out to prompt the user whether to choose to continue writing. The 'Burst' function was developed to prevent the beam current from dropping too fast for the #2 - #4 gyrotrons, which will be discussed in detail in Part 5. When the 'Burst' button is pressed, 'Burst value changed event' will respond. In the burst mode, the 'Burst' button is modified to be red, and we can change the filament power by changing the 'Voltage level', or pressing the 'Up button' or 'Down button'. The color of the 'Burst' button will change back to yellow and the filament power will change back to the original value after the 'Burst duration' has elapsed. Thanks to the use of the producer consumer cycle, the 'Burst' mode does not affect the response and execution of any other events. The program interface is shown in Fig. 9.

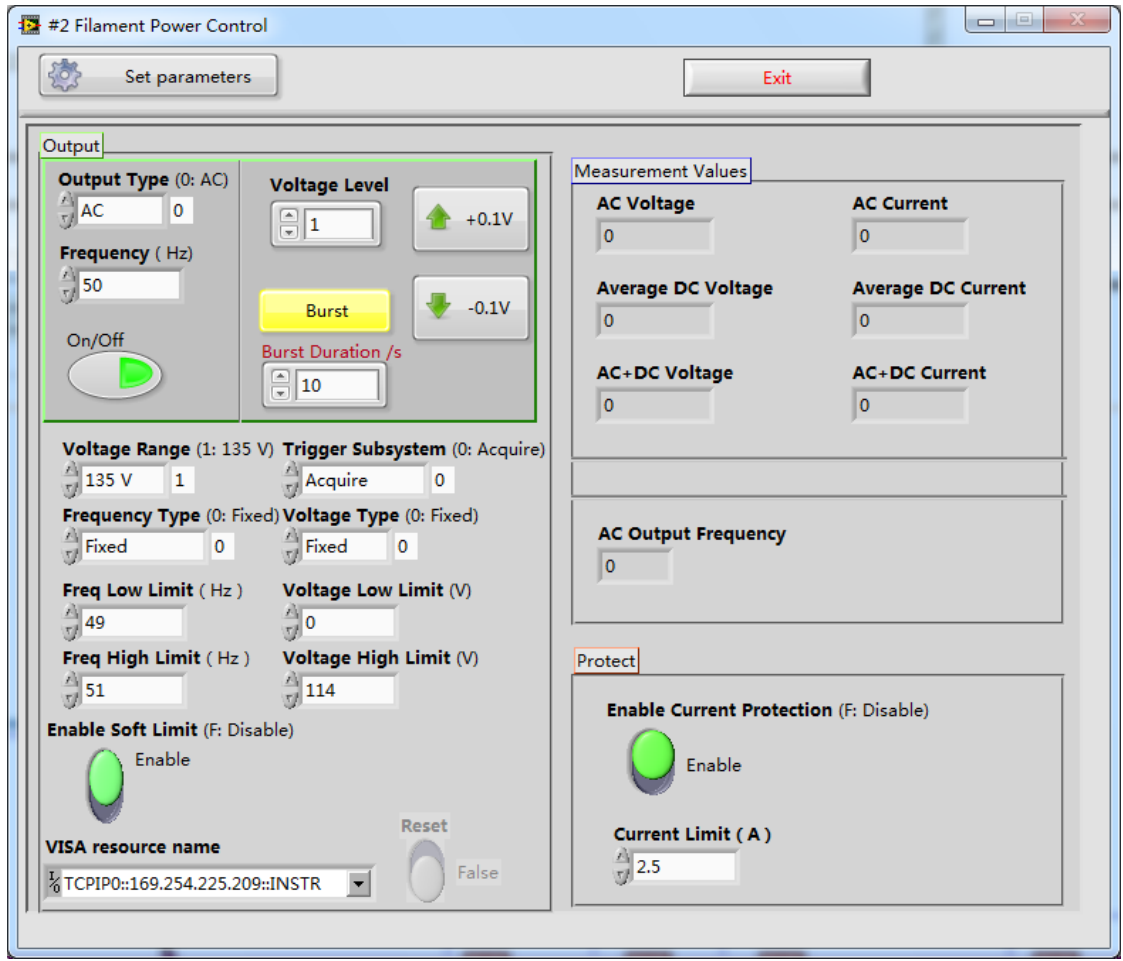

Fig. 9. The program interface for controlling the Keysight 6803A power supply.

## 4. Filament test

Since the filament voltage floats on the cathode high voltage during the operation of the gyrotron, the filament voltage and current during actual gyrotron operation are difficult to measure accurately. In order to obtain the actual filament voltage and current parameters in the gyrotron discharge, we tested the relationship between the primary voltage and current of the filament power supply and the actual voltage and current of the filament in the absence of the cathode voltage. We also tested the stability of the filament power supplies and their control system.

### 4.1. #1 filament test

The filament power test steps can be divided into the following three steps:

(1) Test the isolation transformer, to ensure that the withstand voltage exceeds 70 kV.
(2) Do the test after the filament power supply connected to a dummy load through isolation transformer. After the filament power is connected to the dummy load, gradually increase the filament power, measure the primary and secondary voltage currents, and observe whether the power supply is normal.
(3) Do the test after the filament power supply connected to the gyrotron filament through isolation transformer. This is the most important test, and the test results are shown below.

Firstly, open the auxiliary systems of the gyrotron [9], then turn on the filament power after the main magnetic field reaches 30 A. Test the filament power supply with remote control. The secondary current (the actual filament current) is measured by connecting a 1.896 mΩ resistor in series. The test results are shown in Table 2. All the data recorded in the table are the readings just after the current setting value is adjusted on the control panel. Actually, the current and the voltage on the filament side (secondary) change with the change of the filament resistance. The test results show that the primary power is higher than the secondary power, which means that the CFPS 30/50 power supply (including transformer) consumes some power itself.

Table 2 The test results of the #1 gyrotron filament.

| Primary Power [W] | Primary current [A] | Primary voltage [V] | Secondary current [A] | Secondary voltage [V] |
|---|---|---|---|---|
| 64 | 3.12 | 20.51282 | 7.049345 | |
| 128 | 6.3 | 20.31746 | 9.617321 | 2.6 |
| 192 | 9.18 | 20.91503 | 11.78248 | 4.2 |
| 256 | 11.4 | 22.45614 | 13.2427 | 6 |
| 320 | 13.59 | 23.54673 | 14.95468 | 7.52 |
| 400 | 16.08 | 24.87562 | 17.01913 | 9.56 |
| 480 | 18.42 | 26.05863 | 19.03323 | 11.65 |
| 560 | 20.31 | 27.57262 | 20.74522 | 13.56 |
| 640 | 21.96 | 29.1439 | 22.30614 | 15.59 |
| 704 | 23.04 | 30.55556 | 23.46425 | 17.18 |
| 800 | 24.39 | 32.80033 | 25.02518 | 19.32 |
| 880 | 25.32 | 34.75513 | 26.13293 | 21.01 |
| 960 | 26.19 | 36.65521 | 27.24068 | 22.64 |
| 1040 | 27.06 | 38.43311 | 28.39879 | 24.44 |
| 1120 | 27.87 | 40.18658 | 29.5569 | 26.26 |
| 1200 | 28.32 | 42.37288 | 30.46324 | 27.83 |
| 1264 | 28.47 | 44.39761 | 30.96677 | 28.72 |
| 1312 | 28.62 | 45.84207 | 31.36959 | 29.4 |
| 1344 | 28.74 | 46.76409 | 31.6717 | 29.94 |
| 1376 | 28.86 | 47.67845 | 31.97382 | 30.44 |
| 1392 | 28.98 | 48.03313 | 32.17523 | 30.75 |
| 1408 | 29.07 | 48.43481 | 32.37664 | 31.04 |
| 1424 | 29.13 | 48.88431 | 32.47734 | 31.3 |
| 1416 | 29.1 | 48.65979 | 32.37664 | 31.2 |
| 1400 | 28.95 | 48.35924 | 32.17523 | 30.85 |
| 1392 | 28.92 | 48.13278 | 32.17523 | 30.79 |
| 1384 | 28.86 | 47.95565 | 32.02417 | 30.58 |

The relationship between the primary current and the secondary current is shown in Fig. 10. The filament current can be fitted with the e-exponential function of the primary current displayed by the remote control interface. The certainty coefficient is 0.997. The fitting equation is as follows.

$$I_{filament} = -3.05515 + 9.55202 \cdot \exp(0.04478 \cdot I_{primary}) \tag{1}$$

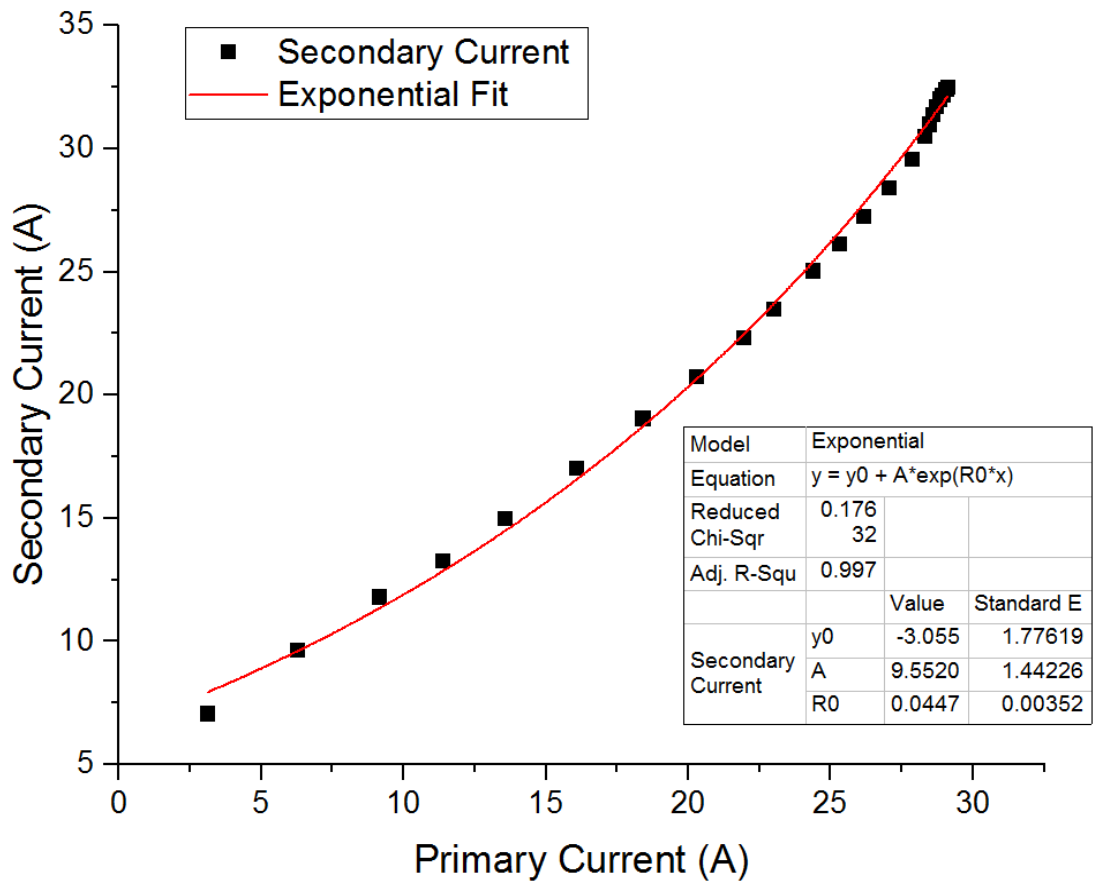

Fig. 10. The relationship between the primary current and the secondary current.

The relationship between the filament current and the filament voltage is shown in Fig. 11. The filament current can be fitted with the e-exponential function of the filament voltage. It can be seen that as the voltage increases, the filament resistance gradually increases.

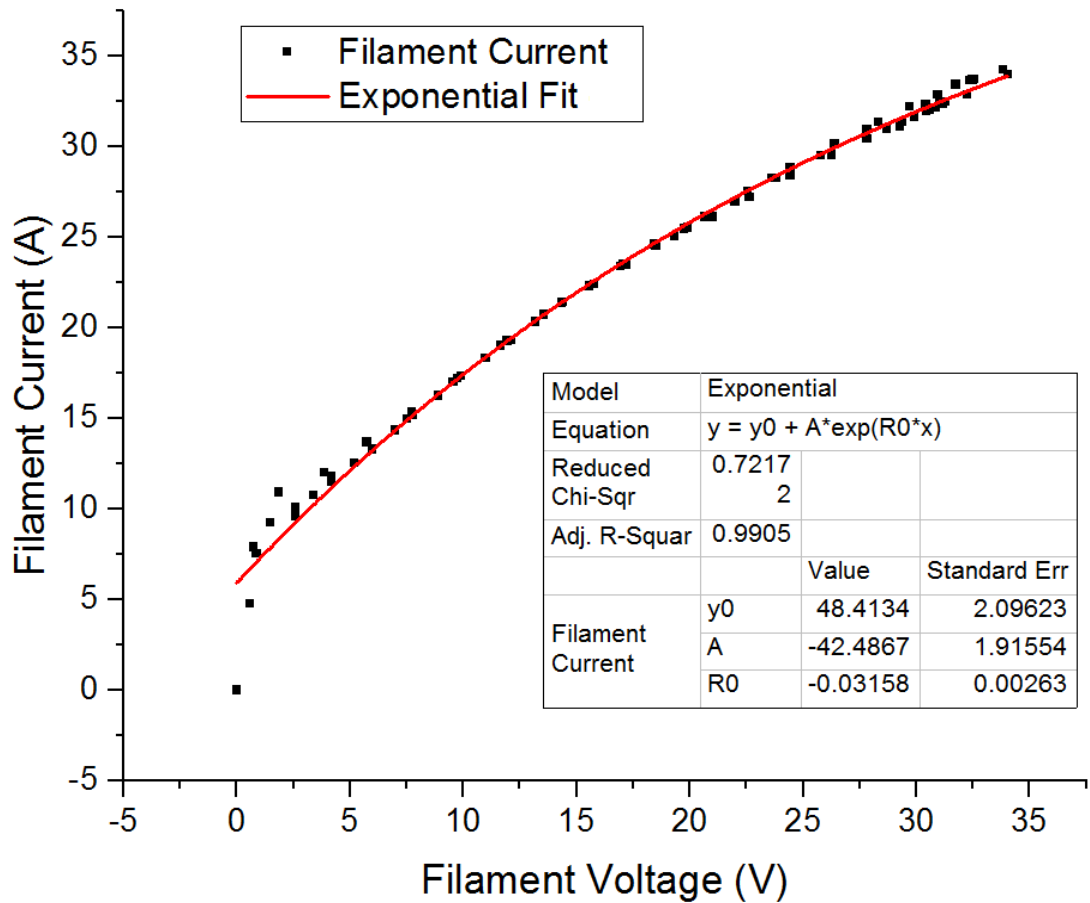

Fig. 11. The relationship between the filament current and the filament voltage of #1 gyrotron.

*4.2.  #2 filament test*

The #2 - #4 gyrotron filament power supply test process is the same as that of the #1 gyrotron filament. It is necessary to test the isolating transformer voltage withstand characteristics first, and then do the load test using the dummy load. Finally, the gyrotron filaments are used as the load to test. Table 3 shows the test results when the Keysight 6803A filament power supply is connected to the #2 gyrotron filament. Since the isolation transformer consumes some power, the primary power is slightly greater than the secondary power, the current transformation ratio is about 3.27, and the voltage transformation ratio is about 3.54. The secondary current is linear with the primary current, and the secondary voltage is also linear with the primary voltage. This is different with that of the #1 CPFS 30/50 power supply. The characteristics of #3 and #4 gyrotron filament power supplies are similar to that of the #2 gyrotron filament power supply and will not be described in detail here.

Table 3 The test results of the #2 gyrotron filament.

| Primary Power [W] | Primary current [Arms] | Primary voltage [Vrms] | Filament Power [W] | Secondary current [Arms] | Secondary voltage [Vrms] | Current ratio | Voltage ratio |
|---|---|---|---|---|---|---|---|
| 0.28 | 0.156 | 1.78 | 0.255 | 0.51 | 0.5 | 3.269 | 3.562 |
| 110.88 | 1.68 | 66 | 99.540 | 5.53 | 18.0 | 3.292 | 3.667 |
| 129.60 | 1.80 | 72 | 121.155 | 5.91 | 20.5 | 3.283 | 3.512 |
| 143.99 | 1.87 | 77 | 133.152 | 6.08 | 21.9 | 3.251 | 3.516 |
| 159.08 | 1.94 | 82 | 147.888 | 6.32 | 23.4 | 3.258 | 3.504 |
| 174.87 | 2.01 | 87 | 161.944 | 6.53 | 24.8 | 3.249 | 3.508 |

The relationship between the filament current and the filament voltage is shown in Fig. 12. The filament current can be fitted with the e-exponential function of the filament voltage. It can be seen that as the voltage increases, the filament resistance gradually increases, which is similar to that of the #1 gyrotron filament.

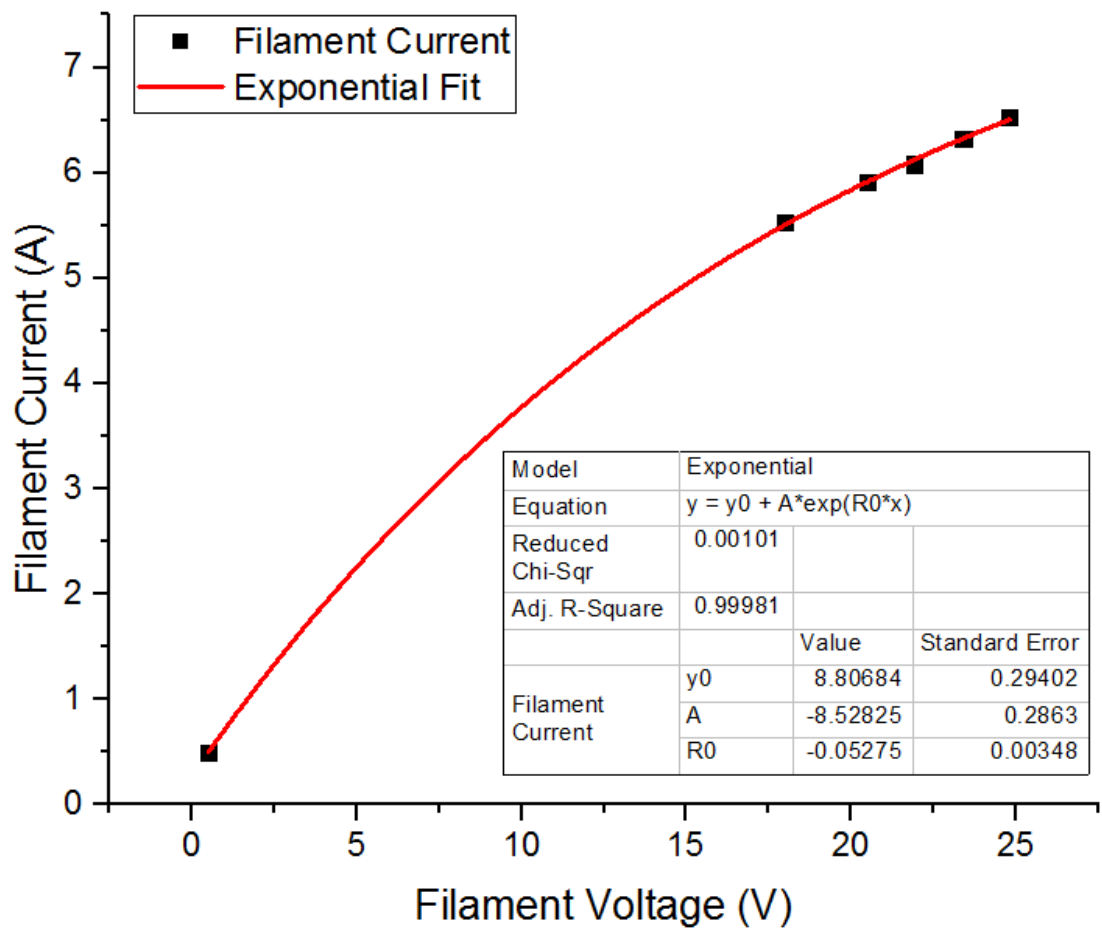

Fig. 12. The relationship between the filament current and the filament voltage of #2 gyrotron.

## 5. Gyrotron test

The first three gyrotrons were tested using the developed filament power control system. In recent experiments, the #1 gyrotron oscillation of 980kW/1s, 903kW/10s, 834kW/95s and 650kW/754s were achieved, the #2 gyrotron oscillation of 721kW/0.5s, 647kW/2s, 499kW/80s and 406kW/98s were achieved, and the #3 gyrotron oscillation of 787kW/20s, 637kW/100s and 559kW/1000s were achieved.

The disruption (i.e., no RF power output due to loss of correct operating mode) may happen because of the cathode current from drooping excessively in the long pulse operation of the gyrotron. Fortunately, the #1 gyrotron has never experienced the disruption caused by cathode current drop. In the test of #2 and #3 gyrotrons, we found that the disruption may happen after a few seconds from the start of the pulse. Therefore, at pulse lengths beyond a few seconds, it is necessary to boost the filament power in order to keep the cathode current from drooping excessively for #2 gyrotron and #3 gyrotrons. The filament voltage will typically be boosted by 1 – 15 % 2 – 10 seconds prior to the beginning of the cathode voltage pulse or 2 - 10 seconds after the beginning of the cathode voltage pulse. The filament voltage need to be returned to its un-boosted value at the end of the cathode pulse. We developed the boost-hold-unboost (we call it 'burst') mode in the filament control system for easy operation. Around the beginning of the cathode voltage pulse, we can click the 'Burst' button on the interface shown in Part 3. The filament voltage will increase according to the set value, and the filament power will change back to the original value after the set duration has elapsed. The 'burst' function can prevent us from forgetting to reduce the filament power after the end of the pulse.

The key signals of #3 gyrotron when using boost and not using boost are shown in Fig. 13. The two shots were made when the cathode voltage is -44.5 kV, the body voltage is 24.5 kV, the superconducting magnet current is 88.4 A, and the initial filament power is 1252 W (119 V, 10.52 A). In the short pulse case, we did not use the boost, the pulse was cut off by the disruption in 30.6 seconds. In the long pulse case, we do use the boost by increasing the filament power to 1288 W (120.7 V, 10.67 A), the pulse ended normally according to the set duration time (100 seconds). We can see that in the absence of the boost, the beam current drops faster, resulting in the disruption.

The gyrotron beam current can be kept stable by controlling the filament heating power. We have tried to develop a PID control algorithm to realize the expected beam current stabilization. However, it may be difficult due to the slow effect of filament power on beam current. In particular, considering the safety of the gyrotron, we have not done relevant experiments. The beam current stabilization control experiments may be done in the next few years.

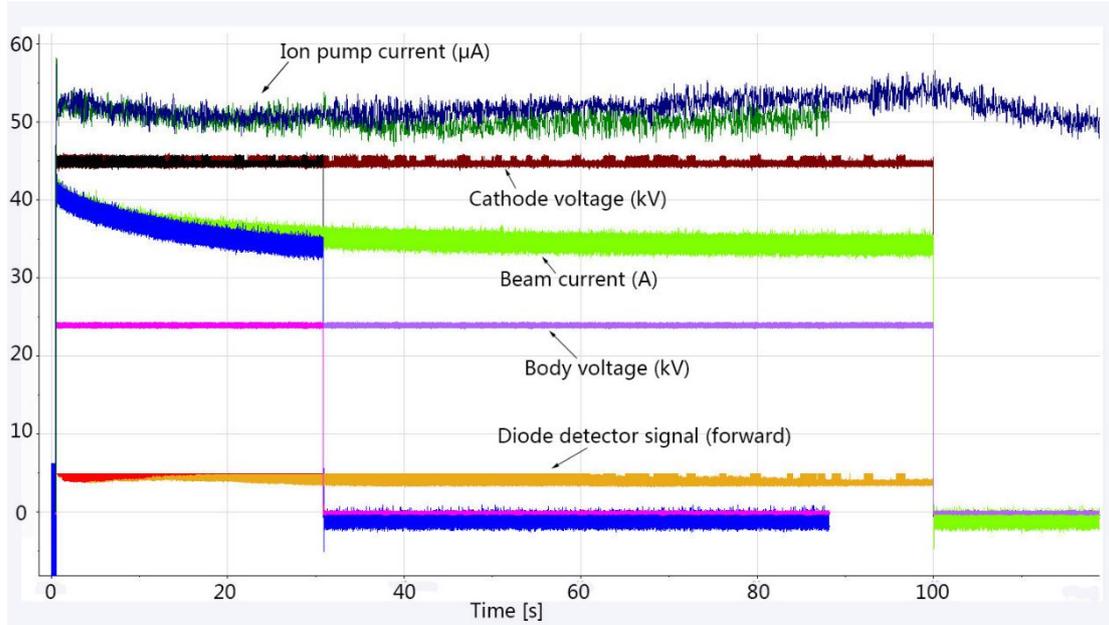

Fig. 13. The key signals of #3 gyrotron with boost and without boost. The short pulse signals show the case in the absence of the boost; the long pulse signals show the case when the boost was used. Except for the boost, the other control parameters of the two shots are the same (the cathode voltage is -44.5 kV, the body voltage is 24.5 kV, the superconducting magnet current is 88.4 A, and the initial filament power is 1252 W, the set full pulse duration is 100s).

## 6. Conclusion

The ECRH system for EAST is being developed in the Institute of Plasma Physics, Chinese Academy of Sciences. There are four gyrotrons in the ECRH system. The first three gyrotrons have been commissioned and tested. Each gyrotron has a filament, which is used to heat the cathode. By adjusting the filament, we can control the beam current, thus control the gyrotron output wave power of the gyrotron.

The filament power control and measurement system has been developed. The filament power control system was described in detail in this paper. The CFPS 30/50 power supply was used as the filament power supply for #1 gyrotron. The Keysight 6803A power supplies were used as the filament power supplies for #2 - #4 gyrotrons. We can control and measure the filament power by controlling the corresponding filament power supply. All the control program were developed by Labview. The filament characteristics were tested using the power control system. The test results show that the filament power control system can work well and the filament current can be fitted with the e-exponential function of the filament voltage. All gyrotron filaments have similar characteristics, the filament resistance gradually increases as the filament power increases. The developed filament control system is important for the power control and stable operation of the gyrotrons in EAST ECRH system.

We are currently artificially increasing the filament power remotely for the safety of the gyrotrons. Based on the filament control system shown in this paper, the soft start of filament is easy to achieve in order to reach the operating value of $V_f$ and $I_f$ in set time. The soft start of filament will be realized in the future.

**Acknowledgements**

This work was supported in part by the National Key R&D Program of China (Grant No. 2017YFE0300401) and the National Magnetic Confinement Fusion Science Program of China (Grant No. 2015GB102003 and 2015GB103000). The authors greatly appreciate the experts from GA, CPI, and GYCOM for the cooperation in the development of the ECRH project on EAST.